\documentclass[showpacs,apl,twocolumn,floatfix]{revtex4-1}

\usepackage{graphicx,float,amsmath}
\usepackage{ mathrsfs }
\usepackage{natbib}

\begin{document}
\date{\today}

\title{Ambipolar spin-spin coupling in p$^+$-GaAs.}

\author{F.~Cadiz$^1$}
\author{D.~Paget$^1$}
\author{A. C. H. ~Rowe$^1$}
\author{S.~Arscott$^2$}

\affiliation{%
$^1$Physique de la Mati\`ere Condens\'ee, Ecole Polytechnique, CNRS, 91128 Palaiseau, France}

\affiliation{%
$^2$Institut d'Electronique, de Micro\'electronique et de Nanotechnologie (IEMN), Universit\'e de Lille, CNRS, Avenue Poincar\'e, Cit\'e Scientifique, 59652 Villeneuve d'Ascq, France}

\begin{abstract}
A novel spin-spin coupling mechanism that occurs during the transport of spin-polarized minority electrons in semiconductors is described. Unlike the Coulomb spin drag, this coupling arises from the ambipolar electric field which is created by the differential movement of the photoelectrons and the photoholes. Like the Coulomb spin drag, it is a pure spin coupling that does not affect charge diffusion. Experimentally, the coupling is studied in $p^+$ GaAs using polarized microluminescence. The coupling manifests itself as an excitation power dependent reduction in the spin polarization at the excitation spot \textit{without} any change of the spatially averaged spin polarization. 
\end{abstract}
\pacs{}
\maketitle

Modification of diffusion to include ambipolar effects in charged heterogeneous media is a topic of interest in several fields, including the study of astrophysical objects \cite{fiedler1993} and plasmas \cite{schulze2011}, as well as in semiconductors \cite{smith1978,zhao2009,paget2012}. The case of semiconductors is of interest both for applications \cite{johnston2002} and because electron gases in semiconductors can be spin polarized. The question of the effect of ambipolar coupling on spin polarized carrier diffusion is yet to be addressed \cite{zhao2009}, but is likely to be of importance for any future bipolar semiconductor spintronic device. A variety of spin-spin and/or spin-charge coupling phenomena have been revealed in semiconductors, for example those due to the spin-orbit interaction \cite{wunderlich2010,jungwirth2012}, those due to the Pauli principle \cite{cadiz_prl2013, cadiz2015b}, as well as the Coulomb spin drag \cite{vignale2002, weber2005}. In the latter case it was shown that a coupling between the $+$ and $-$ spins results in a spin diffusion constant whose magnitude is smaller than the charge diffusion constant. 

Here we describe and study a novel spin-spin coupling mechanism of ambipolar origin which yields a spin diffusion constant whose magnitude is \textit{larger} than the charge diffusion constant. This coupling occurs in the presence of a spatially inhomogeneous gas of spin-polarized photoelectrons and of unpolarized, slower diffusing holes. The differential diffusion of $+$ spin electrons and holes creates an internal electric field which acts on both $+$ and $-$ spins thereby coupling them. The same is true for the $-$ spin electrons. A full description of the coupled diffusion equations is given and the effect is experimentally observed in $p^+$ GaAs.

\begin{figure}[htbp]
\includegraphics[clip,width=7 cm] {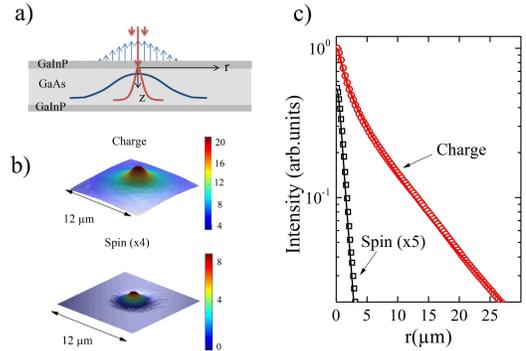}
\caption{(a) The principle of the experimental technique in which the sample is photoexcited by a tightly-focused, circularly-polarized pump (red arrows and lines), and the resulting photoluminescence intensity (blue arrows and lines) and polarization is measured. These two quantities can be used to obtain depth ($z$) integrated charge and spin density profiles as a function of $r$, the radial distance from the excitation spot. (b) Room temperature charge and spin density images obtained from the $p^+$ GaAs sample at low excitation power (0.03 mW) where ambipolar coupling is negligible. (c) The angular averaged profiles obtained from the images in (b). The solid lines are fits using the solution to the uncoupled diffusion equations \cite{favorskiy2010}, from which one obtains the charge ($L$) and spin ($L_s$) diffusion lengths.}
\label{Fig01}
\end{figure}

The sample is a 3 $\mu$m thick, Be-doped ($N_A =  1.5 \times 10^{17}\;\mbox{cm}^{-3}$) GaAs film covered on both sides by passivating GaInP layers which not only reduce the surface recombination velocity, but confine the photocarriers to the active layer. The sample is studied using a microluminescence technique described elsewhere \cite{favorskiy2010} that, as shown in Fig. \ref{Fig01}, creates a spatially inhomogeneous population of spin polarized electrons and of unpolarized holes. This is a pre-requisite for the observation of ambipolar coupling phenomena. The photoexcitation is achieved using a tightly-focused circularly-polarized CW pump (1/e half width of $w =0.6\; \mu$m, energy $1.59$ eV) so that, at the chosen value of $N_A$, ambipolar coupling becomes important for experimentally accessible pump powers. All experiments reported here are at 300 K where other coupling phenomena \cite{weber2005, cadiz_prl2013, cadiz2015b} are negligible. The luminescence intensity and polarization are monitored as a function of distance, $r$, from the excitation spot, from which depth integrated profiles of the photoelectron charge density $n = n_+ + n_-$, and the photoelectron spin density $s = n_+ - n_-$, can be obtained respectively \cite{favorskiy2010}. Here $n_{\pm}$ are the concentrations of electrons of spin $\pm$ with a quantization axis chosen along the direction of light propagation.

\begin{figure}[htbp]
\includegraphics[clip,width=7 cm] {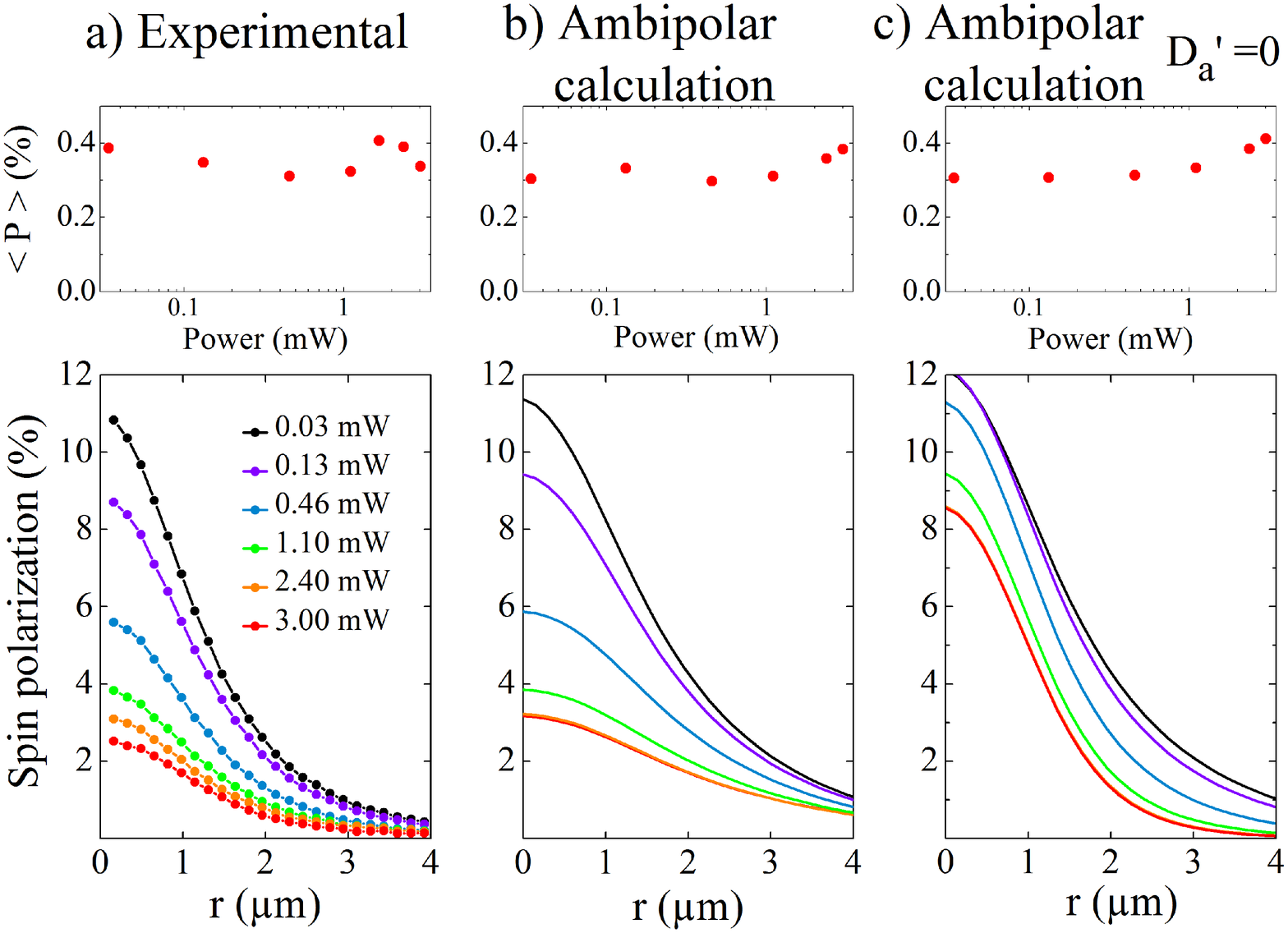}
\caption{The spatial profiles of the spin polarization $\mathscr{P}=s/n$ with increasing values of the excitation power along with the corresponding power dependence of the spatially-averaged polarization $<\mathscr{P}>=<s>/<n>$ (upper inset) shown as obtained experimentally (a), calculated using Eq. \ref{ambiel} which includes the spin-spin coupling (b), and calculated using Eq. \ref{ambisp2} which does not include this coupling. The experimental data reveal a decrease in $\mathscr{P}$ with increasing excitation power at $r = 0$ by about a factor of $6$ \textit{without} any significant change in $<\mathscr{P}>$. A comparison of the experimental and calculated curves reveals that the large reduction in $\mathscr{P}(r=0)$ is a manifestation of the ambipolar spin-spin coupling.}
\label{Fig02}
\end{figure}

The charge and spin density profiles at low power (0.03 mW) illustrate the unipolar regime. These profiles, shown in Fig. \ref{Fig01}(c), are analyzed using the uncoupled diffusion equations \cite{favorskiy2010} (solid lines in the figure), from which the charge diffusion length $L=\sqrt{D_e \tau} =10$ $\mu$m and the spin diffusion length  $L_s=\sqrt{D_e \tau_s} = 0.95$ $\mu$m are obtained. The spatially averaged spin polarization, defined as $<\mathscr{P}>=<s>/<n>$, does not depend on diffusion and is equal, in a two-dimensional picture, to  $\mathscr{P}_i^*(L_s/L)^2$ where $\mathscr{P}_i^*$ is the effective initial polarization including possible losses during thermalization or during diffusion along the $z$ direction. The experimental values of  $<\mathscr{P}> \approx 0.4\%$ shown in Fig. \ref{Fig02}(a), along with the values of $L$ and $L_s$, imply $\mathscr{P}_i^*=0.4$, slightly smaller than its value of 0.5 without losses \cite{meier1984}. The transport parameters for the sample are then characterized by assuming an electron mobility of $\mu_e= 3350 $ cm$^2$/Vs for this doping level \cite{furuta1990,ito1989}, from which the charge diffusion coefficient, $D_e=86$ cm$^2$/s, is obtained using the Einstein relations. Combining this with the measured value of $L$, a minority carrier lifetime $\tau = L^2/D_e = 11.6$ ns is found, close to that measured in similarly doped GaAs \cite{paget2012}. Assuming that in the unipolar limit the spin diffuses with the same diffusion coefficient as the charge, the value of $L_s$ implies a spin relaxation time of $T_1 = 105$ ps. The hole mobility is assumed to be $\mu_h= 220$ cm$^2$/Vs \cite{lowney1991}. As will be seen below, these are all the parameters necessary to describe the spin-spin coupling observed at higher excitation powers.

\begin{figure}[htbp]
\includegraphics[clip,width=7 cm] {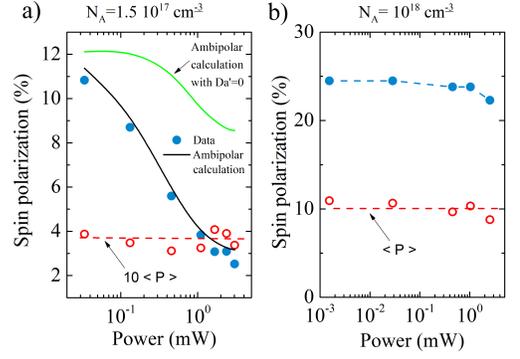}
\caption{(a) The measured polarization at $r=0$ along with the spatially averaged polarization $<\mathscr{P}>$ as a function of excitation power in the GaAs sample doped at $N_A=1.5\times 10^{17}$ cm$^{-3}$. Also shown are the polarizations at $r=0$  calculated with (black line) and without (green line) spin-spin coupling. (b) The same data shown for a GaAs sample with $N_A= 10^{18}\;\mbox{cm}^{-3}$, for which ambipolar effects are strongly reduced and the spin-spin coupling is absent.}
\label{Fig03}
\end{figure}

Fig. \ref{Fig02}(a) shows the spin polarization profiles for increasing excitation powers. At low power, one has  $\mathscr{P}(r=0) = 11\%$ which is a factor $25$ larger than $<\mathscr{P}>$ because the effective lifetime at $r=0$ is not $\tau$ but a diffusion time ($\tau _{diff} \approx w^2/4D_e$). Since $\tau_{diff} \approx 10$ ps $\ll \tau$, this reduces polarization losses by relaxation at $r=0$. Upon increasing the pump power, the polarization at $r=0$ decreases by almost a factor of 6 to $\approx 2$\% at the maximum accessible power (3 mW). It is important to note that this decrease is not due to a decrease in $T_1$ since the spatially-averaged polarization does not change with excitation power (see upper panel of Fig. \ref{Fig02}(a)).

In order to interpret the experimental results it is necessary to calculate the internal ambipolar fields, $\vec E_{\pm}$, created by diffusion of electrons of $\pm$ spins. Neglecting the effects of Pauli blockade, thermoelectric phenomena, and spin Coulomb drag which is screened by the majority holes \cite{cadiz2015b}, the conservation equations for $+$ spins and for holes are: 
\begin{equation}    g_{+}- n_{+}/\tau - (n_{+}- n_{-})/(2T_1) + \vec \nabla \cdot [\sigma_{+} \vec E/q   + D_e \vec \nabla n_{+} ]=  0
\label{electrons} 
\end{equation} 
\begin{equation}    (g_++g_-)-\delta p/\tau + \vec \nabla \cdot [- \sigma_{h} \vec E/q + D_h \vec \nabla \delta p ]=  0,
 \label{holes} \end{equation}

\noindent
where the conservation equation for $-$ spins is obtained by exchanging $+$ and $-$ in Eq. \ref{electrons}.  The generation rate $g_{\pm}$ of $\pm$ spins is strongly peaked at $r=0$ since a tightly focused light excitation is used. Here, $\delta p$ is the photohole concentration $\vec E$ is the internal ambipolar electric field, $q$ is the absolute value of the electron charge and $D_h$ is the hole diffusion constant. The spin conductivities are given by $\sigma_{\pm}=q\mu_e n_{\pm}$ and the electron and hole conductivities are respectively, $\sigma_{e}=q\mu_e n$ and $\sigma_{h}=q\mu_h (N_A^- +\delta p)$, where $N_A^-$ is the concentration of charged acceptors. Calculation of the electric field will assume a 2-dimensional picture (to be justified below) in which the divergences of Eq. \ref{electrons} and Eq. \ref{holes} are approximated by derivatives in the sample plane. Combination of Eq. \ref{electrons} for $+$ spins with the same equation for $-$ spins and with Eq. \ref{holes} for holes shows that $\vec E$ can be written as the sum of three contributions:
\begin{equation}    
\vec E = q\frac{D_h (\vec \nabla \delta p -  \vec \nabla n)}{\sigma_{e} +\sigma_{h}} + \vec E_{+} + \vec E_{-}.
 \label{electricgen} 
 \end{equation}
The first term is caused by a possible disruption of local charge neutrality ($\delta p \neq n$) and will be assumed, as verified below, to be negligible. The contribution  $E_{+}$ is given by:    
\begin{equation}    
\vec E_{+}= q\frac{D_h - D_{e} }{\sigma_{e} +\sigma_{h}} \vec \nabla n_{+}.
 \label{electric} 
 \end{equation}
This field is proportional to the difference in the diffusivities of electrons and holes and is identified as the ambipolar field generated by diffusion of $+$ spins. A similar expression is obtained for $\vec E_{-}$ that is associated with $-$ spins. 

The current $\vec J_+ $ of spins $+$ is then the sum of the diffusion current and of the drift current in  $\vec E_+ +\vec E_-$. The spin-spin coupling is explicit since for example $\vec J_+ $ can be decomposed into two components, $\vec J_{++}$ and $\vec J_{+-}$, proportional to $\nabla n_{+}$ and to $\nabla n_{-}$, respectively. This current, together with  $\vec J_- $, is given by
\begin{equation} 
\begin{pmatrix}
\vec J_+\\
\vec J_-
\end{pmatrix}
=q
\begin{pmatrix}
D_a^{+ +}& D_a ^{+ -}\\
D_a ^{-+}&D_a^{- -}
\end{pmatrix}
\begin{pmatrix}
\vec \nabla n_{+}\\
\vec \nabla n_{-}
\end{pmatrix}
\label{matrix} 
 \end{equation}
where $(\sigma_{e} + \sigma_{h})D_a^{+ +}= \sigma_{+} D_h +(\sigma_{h} +\sigma_{-}) D_e$  and the non diagonal   element, given by $(\sigma_{e} + \sigma_{h})D_a ^{+ -}= \sigma_{+} (D_h - D_e)$ is zero in the unipolar case where $D_e = D_h$. The other matrix elements are obtained by exchanging $+$ and $-$.  The form of the diffusion matrix is similar to that describing the Coulomb spin drag \cite{vignale2002}, with the notable difference that the nondiagonal elements are negative since an outward diffusion of $+$ spins generates an inward flux of $-$ spins. 

It is pointed out that, for a nonzero spin polarization, the two nondiagonal elements of the diffusion matrix are not equal. However, the corresponding diffusion currents,    
\begin{equation}    
\vec J_{+-} = q(D_h - D_{e})\frac{\sigma_{+}\vec \nabla n_{-} }{\sigma_{e} +\sigma_{h}}.
 \label{current} 
 \end{equation}\\
and $J_{-+}$ (obtained by exchanging $+$ and $-$) are equal to first order. Indeed, for nondegenerate electrons, the diffusion length is the same for the two types of spins, so that $\vec \nabla  n_{+}/\vec \nabla  n_{-} \approx n_{+}/ n_{-} $. The currents $\vec J_{+-}$ and $\vec J_{-+}$ describe the flow of comparable numbers of $+$ and $-$ spins per unit time towards the excitation spot which therefore reduces the polarization at $r=0$. This is indeed what is observed experimentally.

Further insight into the nature of the ambipolar spin-spin coupling is gained by studying the diffusion equations for $n$ and $s$, obtained from  Eq. \ref{electrons} by replacing $\vec E$ by its value defined in Eq. \ref{electricgen}. It is immediately clear that the coupling described by the off-diagonal terms of Eq. \ref{matrix} is a pure spin effect, since the charge diffusion equation becomes: 
\begin{equation}    (g_++g_-)- n/\tau + \vec \nabla \cdot [D_{a} \vec \nabla n ] =  0,
\label{ambiel} \end{equation}
where the unipolar diffusion constant is replaced by the usual ambipolar diffusion one \cite{smith1978}, $D_a$ defined by  $(\sigma_{e}+ \sigma_{h})D_a= (\sigma_{e}  D_h +\sigma_{h} D_e)$, without any coupling to the electronic spins. On the other hand spin-spin coupling modifies the spin conservation equation which becomes: 
\begin{equation}   (g_+-g_-)- s/\tau_s + \vec \nabla \cdot [(D_{a}-D_{a}') \vec \nabla s  + D_{a}'  \mathscr{P} \vec \nabla n ]= 0,
\label{ambisp2} 
\end{equation}
where 
 \begin{equation}    
D _a'= D_a^{+- }+ D_a^{-+ }= \frac{\sigma_{e} (D_h - D_e)}{\sigma_{e} + \sigma_{h}}
 \label{D'a} 
 \end{equation} 

\noindent
If spin relaxation is negligible the electronic spin polarization is spatially constant (implying $ \mathscr{P} \vec \nabla n = \vec \nabla s$) and the divergence term in Eq. \ref{ambisp2} reduces to $\vec \nabla \cdot [ D_{a} \vec \nabla s ]$. Spin then diffuses in the same way as charge. In the opposite case where $\mathscr{P}$ decreases with distance as is generally found for local light excitation \cite{favorskiy2010}, the divergence term of Eq. \ref{ambisp2} is of the form $\vec \nabla \cdot [(D_{a}-D_{a}') (\vec \nabla s  + \delta  \mathscr{P} \vec \nabla n)],$ where  $\delta=D_{a}'/D_e$ is close to -1 at high power. Two important conclusions are to be drawn from this analysis. Firstly, the spin diffusion constant $D_{a}-D_{a}'$ is now larger than the charge diffusion constant which is in direct contrast with Coulomb spin drag. Secondly, spin diffusion now depends on charge due to the $\delta  \mathscr{P} \vec \nabla n$ term. This term has the same form as that induced by Coulomb spin drag, or by diffusion of degenerate spins, with the notable difference that in these two cases $\delta > 0$ \cite{cadiz_prl2013, cadiz2015b}.

The agreement between the above model and the experimental results is now verified quantitatively using a numerical resolution of Eq. \ref{ambiel} and Eq. \ref{ambisp2} with the sample parameter values determined above. As shown in Fig. \ref{Fig03}(a), in a low power unipolar regime one finds a polarization at $r=0$ of $11.5 \%$, very close to the measured value. Using $D _a'=0$ (no spin-spin coupling), one finds a slight decrease of polarization at $r=0$ with increasing power which is not sufficient to explain the experimental observations. This decrease is mostly due to the decrease of the ambipolar diffusion constant $D_a$ which results in an increase of the diffusion time $\tau _{diff}$. On the other hand, the inclusion of the spin-spin coupling term, as shown in Fig. \ref{Fig02}(b) accounts very well for the experimental results. It is finally verified that Pauli blockade \cite{cadiz2015b} does not play a role here; for the maximum power the depth averaged value of $n(r=0)$ is calculated to be of the order of $n= 1.6 \times 10^{17}\;\mbox{cm}^{-3}$, which is smaller than the effective density of states of the conduction band at 300 K.

The comparison between experiment and theory is summarized in Fig. \ref{Fig03}(a) which shows excellent agreement with the experimentally measured power dependence of the spin polarization at $r=0$, and the poor one obtained if $D'_a=0$. The ambipolar nature of the experimental results is further confirmed by measurements at 300 K using another sample with an increased acceptor doping of $N_A= 10^{18}\;\mbox{cm}^{-3}$. At this doping density ambipolar coupling is strongly reduced because of the increased majority hole conductivity, so that $D_a \approx D_e$ and $-D'_a \ll D_e$. Indeed, as shown in Fig. \ref{Fig03}(b), the polarization at the excitation spot does not exhibit any decrease as a function of excitation power.
 
\begin{figure}[htbp]
\includegraphics[clip,width=5 cm] {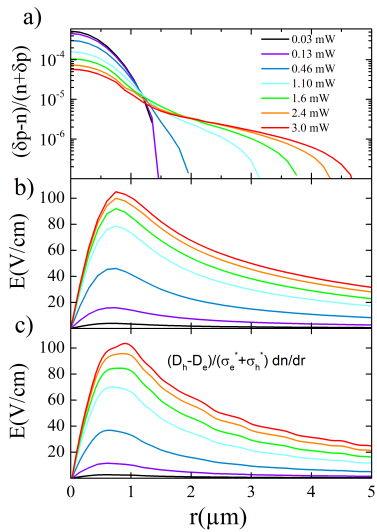}
\caption{(a) Spatial dependence of the relative difference between the photoelectron and photohole density for increasing values of the excitation power. (b) Internal ambipolar electric field. In spite of the small relative difference between the local electron and hole charge densities, this field can be quite large near $r = 0$ at high excitation power. (c) The electric field obtained from the approximate expression Eq. \ref{electric} is close to the numerically calculated result in (b).}
\label{Fig04}
\end{figure}

The two main approximations -- local charge neutrality and the two-dimensional nature of the diffusive transport -- are now justified. The hypothesis of local charge neutrality is verified by removing the approximation $n=\delta p$ and by performing a numerical resolution of Eq. \ref{electrons} and Eq. \ref{holes}, together with Poisson's equation $\vec \nabla \cdot \vec E= (q/\epsilon \epsilon _0) (\delta p -n)$, where $\epsilon _0$ is the vacuum permittivity and $\epsilon$ is the dielectric constant of GaAs. As shown in Fig. \ref{Fig04}(a), it is found that the relative photoinduced electric charge $\left|(n-\delta p)/(n+\delta p)\right|$ is always smaller than $10^{-3}$. As shown in Fig. \ref{Fig04}(b), the internal electric field is peaked near $r = 1$ $\mu$m and can be as large $100$ V/cm at high power. This corresponds to a drift length of several tens of $\mu$m i.e. larger than the charge diffusion length and implies that near $r=0$ charge and spin drift in the internal electric field prevails over diffusion. This is the origin of the large polarization reduction at high excitation powers. Fig. \ref{Fig04}(c) presents the  spatial dependence of the electric field obtained from the approximate expression Eq. \ref{electric}, and it also gives values similar to those of the more general calculation.
 
Note finally that, because the spin-spin coupling does not affect the spatially-averaged polarization $<\mathscr{P}>$, the currents $J_{-+}$ and $J_{+-}$ defined in Eq. \ref{current} should also generate an \textit{increase} of $\mathscr{P}$ at some value of $r$. This increase is not observed in the sample studied here because the magnitude of the spin-spin coupling is proportional to the polarization itself, which is small. A decrease of charge lifetime or an increase of spin lifetime will increase $\mathscr{P}$ and should reveal an absolute maximum in the polarization at some distance from $r=0$. As shown in the supplementary information, the maximum polarization may even exceed $\mathscr{P}_i$.

In conclusion, it has been shown both theoretically and experimentally that minority electron spin diffusion in the presence of slower diffusing photoholes generates a coupling between electon spins $+$ and $-$ such that the outward flow of spins $\pm$ generates an inward flow of spins $\mp$. This is a pure spin coupling which does not affect charge diffusion. The diffusion constant is then described by a matrix with negative nondiagonal elements and increased values of the diagonal elements with respect to the unipolar regime. This effect strongly reduces the spin polarization at the excitation spot for excitation powers that are sufficiently high to ensure that the inward ambipolar spin currents are comparable with the outward diffusive currents. One of us (F. C.) is grateful to CONICYT Grant Becas Chile for supporting his work.

\bibliographystyle{apsrev}

%\bibliography{cadizref}

\end{document}